\documentclass[a4paper,11pt]{article}
\usepackage{jinstpub}

\usepackage{graphicx}
\usepackage{amsmath}

\title{Process Development and First Cryogenic Operation of Compact Germanium Ring-Contact HPGe Prototypes}

\author[a,1]{Kunming Dong%
\note{These authors contributed equally to this work.}}
\author[a,1]{Shasika Panamaldeniya}
\author[a]{Dongming Mei}

\affiliation[a]{Department of Physics, University of South Dakota,
           Vermillion, SD 57069, USA}

\emailAdd{dongming.mei@usd.edu}

\abstract{Rare-event experiments such as LEGEND-1000 pursue a ton-scale search for neutrinoless double-beta decay in \textsuperscript{76}Ge and therefore require HPGe detectors with excellent energy resolution, low electronic noise, and scalable low-background packaging. Point-contact-style electrodes have enabled powerful pulse-shape discrimination in prior \textsuperscript{76}Ge programs, while inverted-coaxial point-contact (ICPC) detectors extend these advantages to multi-kilogram crystals and form the current LEGEND baseline. Further increasing the mass per detector remains attractive because it can reduce channel count, cabling complexity, and total passive surface area.

The germanium ring-contact (GeRC) concept addresses this scaling problem through a recessed ring-and-groove electrode topology designed to preserve point-contact-like low-capacitance signal formation in larger crystals. However, reliable GeRC fabrication has not yet been demonstrated, largely because the non-planar groove surfaces complicate machining, surface recovery, conformal passivation, and especially the eventual formation of a robust lithium-diffused outer contact.

This work reports the fabrication and first cryogenic operation of two compact n-type GeRC process-validation prototypes produced from in-house HPGe crystals at the University of South Dakota. An optimized workflow for core drilling, groove cutting, non-planar polishing, conformal a-Ge encapsulation, Al patterning, and GeRC-specific cryogenic mounting was implemented. Two independent sputtering systems were used to test whether the thin-film sequence remains operable across substantially different deposition environments. At 77 K, both devices could be biased stably, showed an inferred depletion onset near 340 V from a pulser-based capacitance proxy consistent with electrostatic modeling, and produced identifiable full-energy peaks from \(^{241}\mathrm{Am}\) and \(^{137}\mathrm{Cs}\). These results establish a proof-of-principle process and readout baseline for the geometry-specific steps of GeRC development. They do not yet constitute a deployment-ready large-mass GeRC technology; rather, they define the foundation for the next step, namely integration of conformal lithium-paint deposition and controlled diffusion on the ring-and-groove topology.}


\begin{document}
\maketitle
\flushbottom

\section{Background}

High-purity germanium detectors have evolved substantially over the past decades to meet the demanding requirements of rare-event physics~\cite{DAndrea2021_Universe_GeReview,Ackermann2013_EPJC_GERDA,Knoll2010_RDM}. Coaxial HPGe detectors, traditionally used in $\gamma$-ray spectroscopy and early neutrinoless double-beta decay (0$\nu\beta\beta$) experiments, provided large crystal masses but with relatively high capacitance and limited event discrimination capability~\cite{Ackermann2013_EPJC_GERDA,DAndrea2021_Universe_GeReview,Knoll2010_RDM}. The development of point-contact detector geometries addressed these limitations. In particular, p-type point-contact (PPC) detectors and Broad Energy Ge (BEGe) detectors were introduced in the 2000s to achieve low capacitance and strong pulse-shape discrimination (PSD)~\cite{Luke1989_IEEE_ShapedField,Barbeau2007_JCAP_PPC,Agostini2015_EPJC_BEGeProd,Agostini2019_EPJC_BEGeChar}. Experiments like the Majorana Demonstrator (using PPCs) and GERDA Phase II (using BEGes) showed that such detectors could effectively distinguish single-site signal events from multi-site background interactions, dramatically improving background rejection in \textsuperscript{76}Ge 0$\nu\beta\beta$ searches~\cite{Alvis2019_MJD_MSE,Agostini2013_GERDA_PSD,Agostini2022_EPJC_GERDAPSA,Arnquist2023_MJD_Final}. However, these PPC and BEGe devices typically have masses on the order of 0.5--1 kg each; scaling a strict point-contact design to substantially larger crystals becomes challenging because ensuring full depletion with acceptable bias voltages can become impractical~\cite{Agostini2019_EPJC_BEGeChar,Comellato2021_EPJC_Collective,Agostini2021_EPJC_ICPC}. Above roughly the 1--2 kg scale, a simple point-contact geometry can suffer from regions of low field (partial depletion or ``pinch-off'') that degrade charge collection and PSD performance~\cite{Bonet2021_EPJC_CONUS,Jany2021_EPJC_Prototype,Agostini2021_EPJC_ICPC}. This practical size limitation motivated new detector designs to maintain high performance in larger germanium masses~\cite{Agostini2021_EPJC_ICPC,Abgrall2021_LEGEND1000,DAndrea2021_Universe_GeReview}.

The inverted coaxial point-contact (ICPC) geometry was developed to enable p-type crystals in the 2--4 kg range while retaining the advantages of a small readout electrode~\cite{Cooper2012_NIMA_HPGeTrackingImaging,Agostini2021_EPJC_ICPC,Abgrall2021_LEGEND1000}. The ICPC design, originally conceived at ORNL, introduces an extended coaxial-like geometry with a central bore and an inverted electrode configuration that shapes the electric field more uniformly throughout a large crystal~\cite{Cooper2012_NIMA_HPGeTrackingImaging,Agostini2021_EPJC_ICPC,Abgrall2021_LEGEND1000}. By preserving a point-contact anode and a coaxial cylindrical geometry, ICPC detectors achieve low capacitance and excellent PSD in devices considerably larger than standard PPCs~\cite{Agostini2021_EPJC_ICPC,Abgrall2021_LEGEND1000,Agostini2022_EPJC_GERDAPSA}. This technology has been adopted in current 0$\nu\beta\beta$ projects: for example, the LEGEND-200 experiment operates modules of enriched-Ge ICPC detectors, and the upcoming LEGEND-1000 is planned to deploy on the order of 400 ICPC units of 2--4 kg each to reach a total active mass of one tonne~\cite{Saleh2026_arXiv_LEGEND200,Agostini2021_EPJC_ICPC,Abgrall2021_LEGEND1000}. Using fewer, larger detectors is highly beneficial for ultra-low-background experiments. It reduces the number of electronic channels and cryostat penetrations, simplifies mechanical integration, and decreases the total surface area that could host radioactive contaminants. Each of these factors contributes to a lower background footprint per unit mass of detector~\cite{Agostini2021_EPJC_ICPC,Abgrall2021_LEGEND1000,DAndrea2021_Universe_GeReview}. Nonetheless, even the ICPC design may encounter limitations as one pushes to ever larger crystals. It remains to be demonstrated that an ICPC below \textasciitilde{}4.5 kg can be fully depleted without any undepleted volume; simulations suggest that at \textasciitilde{}5 kg, maintaining adequate field throughout the entire crystal becomes difficult without introducing new electrode structures~\cite{Agostini2021_EPJC_ICPC,Abgrall2021_LEGEND1000,Saleh2026_arXiv_LEGEND200}. Moreover, the primary driver for exploring new HPGe geometries beyond the current ICPC scale is scalability at the module and array level. In a ton-scale \textsuperscript{76}Ge experiment, each detector requires at least one high-voltage line and one readout channel; therefore, increasing the mass per detector directly reduces the number of channels and associated cables, simplifies mechanical integration, and lowers the background contribution from cabling and feedthroughs. In addition, larger crystals generally provide higher containment efficiency and an improved mass-to-surface ratio, which together can reduce the background rate per kilogram and strengthen overall sensitivity. These scaling considerations motivate the new geometry as a route to preserve point-contact-like low-capacitance signal formation while extending single-crystal mass into the multi-kilogram regime~\cite{Abgrall2021_LEGEND1000,Agostini2021_EPJC_ICPC,DAndrea2021_Universe_GeReview}.

\begin{figure}[t]
    \centering
    \includegraphics[width=0.5\linewidth]{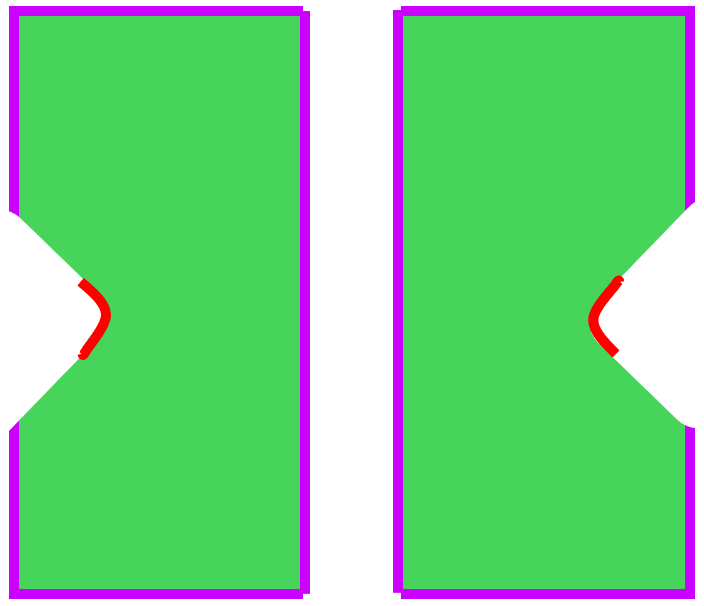}
    \caption{Schematic cross-sectional view of a germanium ring-contact detector geometry. The green region represents the germanium bulk. The purple boundary denotes the outer detector surface, which serves as the large-area opposing electrode in a p-type implementation. The recessed features visible on the left and right sides correspond to the two intersections of a single annular groove with the cross-sectional plane. The red segments indicate the corresponding cross-sectional intersections of the continuous ring-shaped p$^{+}$ readout contact formed on the recessed groove surface. This geometry preserves a spatially localized readout electrode while increasing the detector volume, thereby supporting low-capacitance operation in larger-mass germanium detectors~\cite{Leone2022_Thesis_RingContact,Hull2022_PHDS_RCD,Hull2024_PHDS_RCD,Radford2024_RingContactTalk,Radford2018_PIRE_RingContact}.}
    \label{fig:ring_contact_schematic}
\end{figure}

A next-generation concept addressing these needs is the germanium ring-contact detector (GeRC)~\cite{Leone2022_Thesis_RingContact,Hull2022_PHDS_RCD,Hull2024_PHDS_RCD,Radford2024_RingContactTalk,Radford2018_PIRE_RingContact}. Proposed by D.~Radford around 2018, the ring-contact geometry introduces a modified small-electrode topology aimed at enabling multi-kilogram p-type Ge detectors while preserving the low-capacitance response characteristic of point-contact devices~\cite{Leone2022_Thesis_RingContact,Hull2022_PHDS_RCD,Hull2024_PHDS_RCD,Radford2024_RingContactTalk,Radford2018_PIRE_RingContact}. In this design, the p$^{+}$ readout electrode is implemented not as a single point on a planar face, but as a narrow ring electrode located on a recessed surface of the crystal~\cite{Leone2022_Thesis_RingContact,Hull2022_PHDS_RCD,Hull2024_PHDS_RCD,Radford2024_RingContactTalk,Radford2018_PIRE_RingContact}. Figure~\ref{fig:ring_contact_schematic} presents a schematic cross-sectional illustration of representative groove-based ring-contact configurations. In such implementations, the ring-shaped p$^{+}$ contact is formed on the recessed annular surface, while the remaining outer surface of the crystal serves as the large-area opposing n$^{+}$ electrode~\cite{Leone2022_Thesis_RingContact,Hull2022_PHDS_RCD,Hull2024_PHDS_RCD,Radford2024_RingContactTalk,Radford2018_PIRE_RingContact}. By relocating the small readout contact from a planar point to a recessed annular surface, the detector geometry modifies both the electric-field distribution and the weighting potential, providing a practical route toward large detector masses with point-contact-like signal formation and correspondingly low electronic capacitance~\cite{Leone2022_Thesis_RingContact,Hull2022_PHDS_RCD,Hull2024_PHDS_RCD,Radford2024_RingContactTalk,Radford2018_PIRE_RingContact}. Essentially, the ring-contact detector is a clever hybrid of point-contact and coaxial philosophies: it maintains a tiny collecting electrode (and hence low capacitance) while using a ring-and-groove geometry to shape the electric field lines and reach deeper into a large crystal~\cite{Leone2022_Thesis_RingContact,Hull2022_PHDS_RCD,Leone2021_UNC_Poster}. Field simulations indicate that this approach can achieve full depletion in significantly larger crystals than conventional ICPCs; masses up to approximately 7 kg per crystal could be fully depleted at bias voltages of only a few kilovolts, comparable to present ICPC operation~\cite{Hull2022_PHDS_RCD,Hull2024_PHDS_RCD,Leone2021_UNC_Poster}. If realized, such detectors would allow further reduction of channel count and detector-unit overhead in LEGEND-1000 and beyond, opening a path to improved sensitivity through higher active mass per channel~\cite{Abgrall2021_LEGEND1000,Hull2024_PHDS_RCD,Dong2024_APS_GeRC}. At present, however, GeRC should still be viewed as a promising but experimentally immature geometry whose central challenges are dominated by fabrication and contact integration rather than by electrostatic design alone. It is important to note that the ring-contact concept is envisioned as a complementary innovation rather than an outright replacement for ICPC technology~\cite{Abgrall2021_LEGEND1000,Agostini2021_EPJC_ICPC,Dong2024_APS_GeRC}. The baseline plan for LEGEND-1000 is still to employ ICPC detectors, which are a proven solution; ring-contact detectors would represent an additional option to deploy even larger crystals or to reduce surface dead layers if they can be developed to a mature state~\cite{Abgrall2021_LEGEND1000,Agostini2021_EPJC_ICPC,Wei2019_JINST_aGeDetectors,Luke1992_aGeBipolar}. In essence, the ring-contact design extends the evolution of HPGe detectors by building on the success of ICPCs and pushing the performance envelope further in terms of size and background reduction~\cite{Abgrall2021_LEGEND1000,Agostini2021_EPJC_ICPC,Hull2024_PHDS_RCD}. Both ICPC and ring-contact detectors share the same goals of large mass, low noise, and excellent PSD, and a future tonne-scale experiment could benefit from a mix of these device types optimized for different operational niches~\cite{Agostini2021_EPJC_ICPC,Abgrall2021_LEGEND1000,Hull2024_PHDS_RCD}.

While the potential advantages of the ring-contact geometry are compelling, fabricating such detectors presents significant technical challenges that are not encountered, or are far less severe, in established PPC and ICPC production~\cite{Dong2024_APS_GeRC,Dong2026_arXiv_HybridProcess}. Beyond contact formation, the mechanical machining burden is substantially higher: the ring-contact topology requires a central bore and recessed ring-and-groove features, which introduce localized stress concentrations and tighter tolerance requirements during cutting and shaping~\cite{Dong2024_APS_GeRC,Dong2026_arXiv_HybridProcess}. Because detector-grade Ge is mechanically brittle and sensitive to thermal and torque-induced stress, machining steps that are routine for simpler geometries can lead to cracking or subsurface damage if heat removal, fixturing, and tool geometry are not carefully controlled~\cite{Amman2020_SegmentedAS,Wei2019_JINST_aGeDetectors,Martin2008_Reliability}. Consequently, reliable GeRC fabrication demands a dedicated machining workflow (e.g., improved slotting/cutting approaches, better thermal management, and higher-torque stability during fixturing), followed by aggressive surface-damage removal through deep etching and chemical cleaning to establish reproducible contact interfaces on core-and-groove features~\cite{Amman2020_SegmentedAS,Wei2019_JINST_aGeDetectors,Martin2008_Reliability,Dong2026_arXiv_HybridProcess}.

A second, and often dominant, risk is the formation of a continuous, low-leakage n+ outer electrode on the recessed topology~\cite{Dong2024_APS_GeRC,Dong2026_arXiv_HybridProcess}. For large ring-contact crystals, uniform one-step lithium evaporation followed by diffusion is intrinsically complicated by the ring-and-groove geometry, because achieving conformal lithium coverage on vertical sidewalls and within recessed grooves is nontrivial~\cite{Dong2024_APS_GeRC,Dong2026_arXiv_HybridProcess}. This has been identified as the principal fabrication bottleneck in the ring-contact development path; correspondingly, reliable production has not yet been demonstrated even though simulated fields and signal formation appear promising~\cite{Leone2022_Thesis_RingContact,Dong2024_APS_GeRC,Dong2026_arXiv_HybridProcess}.

In this context, the process strategy is not to rely on a ``standard'' evaporate-and-diffuse step for GeRC, but instead to employ a conformal lithium deposition route that is inherently compatible with complex surfaces.

Accordingly, the program adopts lithium painting as a process pathfinder specifically aimed at improving conformality on core-and-groove features~\cite{Yin1969_Thesis_LiPaint,Morgan1969_GeLiPlanar,Levy1968_AnnularGeLi,Dong2026_arXiv_HybridProcess}. In this traditional approach, a lithium--oil suspension is applied as a coating and then followed by a controlled thermal diffusion schedule~\cite{Yin1969_Thesis_LiPaint,Morgan1969_GeLiPlanar,Levy1968_AnnularGeLi,Fuller1953_LiDiff_GeSi}. In this method, the lithium--oil mixture enables a more uniform ``paint-on'' deposition, and diffusion uniformity is further improved through segmented diffusion protocols and tighter temperature control~\cite{Fuller1953_LiDiff_GeSi,Pell1957_LiSolub_Ge,Jindal1970_LiGeMaterials,Dong2026_arXiv_HybridProcess}. Consistent with this motivation, the hybrid planar process-vehicle KL01 demonstrates that paint-on deposition followed by controlled diffusion can yield stable depletion and pA-scale leakage at 77 K, and that the lithium-paint n+ contact is compatible with subsequent thin-film steps (a-Ge/Al signal electrode and a-Ge passivation)~\cite{Dong2026_arXiv_HybridProcess,Yang2022_WraparoundLiAGe,Ghosh2025_LiGeContact,Wei2019_JINST_aGeDetectors}.

Despite this progress, transferring lithium painting from planar coupons to a true ring-contact topology still involves unresolved implementation details (e.g., coating uniformity and edge control inside grooves and along vertical walls, diffusion repeatability after core-and-groove polishing/etching, and failure-mode management during subsequent wet processing)~\cite{Dong2024_APS_GeRC,Dong2026_arXiv_HybridProcess}. For this reason, two compact GeRC prototypes were first fabricated using an all--thin-film amorphous-contact approach (a-Ge passivation with a-Ge/Al metallized electrodes)~\cite{Dong2024_APS_GeRC,Panamaldeniya2026_arXiv_ICPCaGe,Luke2000_OrthStrip,Hull2005_aGeContacts,Amman2020_SegmentedAS}. The goal was not to demonstrate the final large-mass GeRC architecture, but to provide an experimental proof-of-principle for the recessed ring topology beyond simulation and to isolate the geometry-specific process risks before introducing lithium diffusion~\cite{Dong2024_APS_GeRC,Panamaldeniya2026_arXiv_ICPCaGe}. The prototypes were also used to check whether the required cutting and core-and-groove polishing introduce bulk damage that would compromise detector operation~\cite{Dong2024_APS_GeRC,Amman2020_SegmentedAS,Wei2019_JINST_aGeDetectors}. In addition, this effort was intended to mature the process steps that are unique to GeRC, most notably defining a narrow, continuous ring-shaped electrode that is fundamentally different from conventional point or planar contacts~\cite{Dong2024_APS_GeRC,Luke2000_OrthStrip,Hull2005_aGeContacts,Amman2020_SegmentedAS}. This staged approach establishes a practical baseline for GeRC fabrication while isolating the remaining lithium-paint transfer risks, thereby enabling a stepwise route toward a lithium-based GeRC~\cite{Dong2024_APS_GeRC,Dong2026_arXiv_HybridProcess}.

Amorphous semiconductor contacts (a-Ge/a-Si) provide clear advantages for small detectors and development studies because they can be thin, conformal, and strongly passivating when properly prepared~\cite{Luke2000_OrthStrip,Hull2005_aGeContacts,Amman2020_SegmentedAS,Panth2020_EPJC_Cryo,Wei2019_JINST_aGeDetectors}. However, their role should be stated conservatively in the ring-contact scaling context. Temperature-dependent leakage analyses commonly indicate that amorphous-Ge blocking barriers are typically smaller (order 0.3 eV) than the effective Li n+--p junction barrier at 77 K (order 0.5 eV), underscoring why the Li-based outer contact remains the more robust high-voltage solution in large-mass HPGe instrumentation~\cite{Wei2020_EPJC_CBH,Panth2022_NIMA_TempBarrier,Dong2026_arXiv_HybridProcess,Haller1981_PhysicsUPGe}. Indeed, lithium-diffused contacts remain the most widely used outer-electrode technology for large HPGe detectors because they provide mechanically durable electrodes and robust high-voltage operation, even though they introduce an inactive/transition layer that must be controlled for optimal charge collection and background performance~\cite{Haller1981_PhysicsUPGe,Ackermann2013_EPJC_GERDA,Agostini2021_EPJC_ICPC,Aguayo2013_NIMA_LiNplus,Andreotti2014_ARI_DeadLayer,Ma2017_ARI_InactiveLayer,Dai2022_arXiv_LiInactive}.

In summary, the demonstrated all--thin-film GeRC prototypes establish a reproducible baseline for the most geometry-specific steps, including core-and-groove machining, surface conditioning, and conformal passivation/contact deposition~\cite{Dong2024_APS_GeRC,Amman2020_SegmentedAS,Wei2019_JINST_aGeDetectors}. In parallel, the KL01 planar vehicle validates the lithium-paint diffusion protocol as a conformal n+ formation strategy intended for ring-and-groove features~\cite{Dong2026_arXiv_HybridProcess,Yin1969_Thesis_LiPaint,Yang2022_WraparoundLiAGe,Ghosh2025_LiGeContact}. The next milestone is to integrate these two validated threads: mature GeRC machining and lithium painting/diffusion~\cite{Dong2024_APS_GeRC,Dong2026_arXiv_HybridProcess}. This integration will enable a lithium-based GeRC device that completes the end-to-end fabrication flow required for scalable ring-contact modules relevant to LEGEND-class instrumentation~\cite{Dong2024_APS_GeRC,Dong2026_arXiv_HybridProcess,Abgrall2021_LEGEND1000}.

\section{GeRC Detector Design and Electrostatic Modeling}

\subsection{Design rationale and contact strategy}
The GeRC concept uses a recessed ring-shaped readout electrode to preserve point-contact-like capacitance while reshaping the internal field for large-volume charge collection~\cite{Luke1989_IEEE_ShapedField,Leone2022_Thesis_RingContact,Hull2024_PHDS_RCD}. In a mature implementation, a lithium-diffused n$^{+}$ outer electrode is attractive because it is mechanically durable and supports robust high-voltage operation~\cite{Haller1981_PhysicsUPGe,Morgan1969_GeLiPlanar,Yang2022_WraparoundLiAGe}. However, lithium diffusion also introduces a comparatively thick dead layer and a transition region whose thickness and charge-collection properties must be controlled~\cite{Aguayo2013_NIMA_LiNplus,Ma2017_ARI_InactiveLayer,Dai2022_arXiv_LiInactive}. In contrast, amorphous-Ge (a-Ge) thin-film contacts can be conformal and strongly passivating when properly prepared, and they add only a thin entrance window~\cite{Luke1992_aGeBipolar,Hull2005_aGeContacts,Panth2020_EPJC_Cryo}. Their blocking barriers are typically smaller and more process-sensitive than lithium-based contacts, which can limit the maximum bias margin in some configurations~\cite{Hull2005_aGeContacts,Wei2020_EPJC_CBH,Panth2022_NIMA_TempBarrier}.

For the present work, lithium diffusion on a core-and-groove GeRC topology is still a development risk because conformal lithium coverage on vertical walls and within recessed grooves has not yet been established in our process. We therefore fabricated two compact GeRC prototypes using a fully encapsulating thin-film approach. This choice isolates the geometry-specific risks and tests whether the required cutting, groove machining, and non-planar polishing introduce bulk damage that could prevent stable detector operation. It also provides an opportunity to evaluate the feasibility of ``thin contacts'' on complex 3D detector surfaces before integrating lithium diffusion in later generations. Accordingly, the present results should be interpreted as a fabrication, biasing, and spectroscopy proof-of-principle for compact process-validation devices rather than as a full demonstration of the ultimate large-mass p-type GeRC architecture.

\subsection{Prototype geometry, electrode layout, and crystal selection}
\begin{figure}[t]
    \centering
    \includegraphics[width=0.5\linewidth]{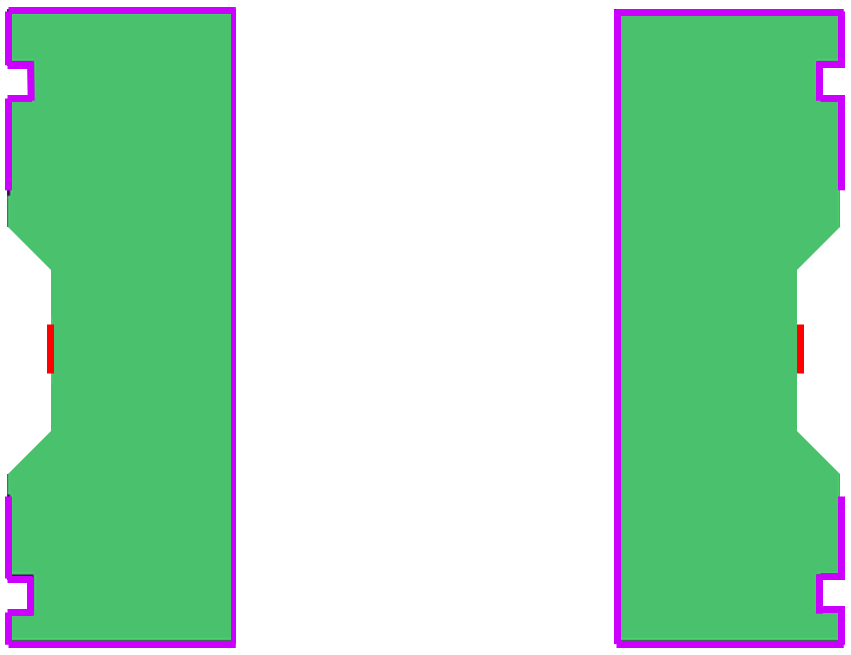}
    \caption{Axial cross-sectional view of the compact GeRC geometry and electrode assignment. The green region represents the HPGe bulk. The white regions correspond to the central bore and the machined annular groove. The red segments denote the two intersections of the continuous annular ring electrode with the axial cutting plane; this ring serves as the small-area signal readout contact. The purple boundary marks the remaining surfaces assigned to the large-area HV electrode. The shallow recessed features near the top and bottom edges are wing-and-groove handling structures introduced to improve mechanical handling during fabrication.}
    \label{fig:gerc_cross_section}
\end{figure}

\begin{figure}[t]
    \centering
    \includegraphics[width=0.4\linewidth]{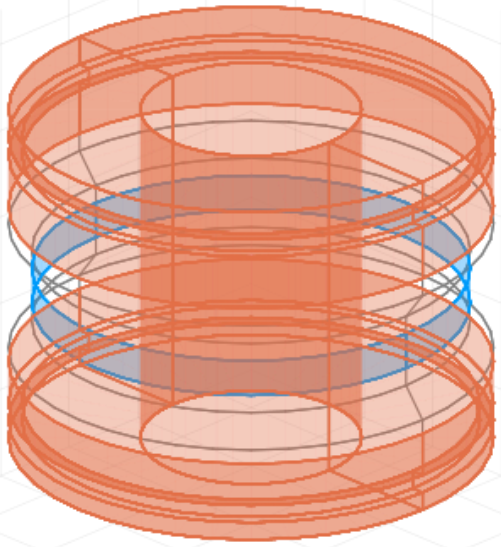}
    \caption{Three-dimensional rendering of the compact GeRC geometry generated using \texttt{SSD.jl}. This view makes the azimuthal continuity of the annular ring electrode explicit and visualizes the full electrode topology used in the detector model. The rendering shows how the localized ring readout contact is embedded within the surrounding large-area HV electrode structure, which includes the bore surface, the top and bottom faces, and the connected outer-sidewall regions above and below the groove.}
    \label{fig:gerc_3d_rendering}
\end{figure}

The compact GeRC crystal is a cylindrical HPGe piece with a height of 19~mm and a diameter of 25~mm. A central bore is machined along the detector axis, and an annular groove is machined on the outer surface. A narrow ring electrode is defined at the center of this groove and serves as the signal readout contact. The detector geometry and electrode assignment used in this work are shown in Fig.~\ref{fig:gerc_cross_section} and Fig.~\ref{fig:gerc_3d_rendering}. The axial cross-sectional view in Fig.~\ref{fig:gerc_cross_section} shows that the ring contact appears as two red segments because the cutting plane intersects a single continuous annular electrode at two locations. The corresponding three-dimensional rendering in Fig.~\ref{fig:gerc_3d_rendering}, generated using \texttt{SolidStateDetectors.jl} (\texttt{SSD.jl}) ~\cite{Abt2021_JINST_SSD}, illustrates the full electrode topology adopted in the detector model. In the baseline SSD.jl model used below, the ring electrode width is 2~mm. The two fabricated prototypes retained ring widths of approximately 0.8~mm (SAP18) and 2~mm (KMRC01), respectively (see Sec.~3.5). The remaining designated electrode regions, including the bore surface, the top and bottom faces, and the connected outer-sidewall regions above and below the groove, together form the large-area high-voltage (HV) electrode.

A key requirement for this electrode topology is a unipolar bulk, meaning that the net impurity concentration must not change sign within the active volume~\cite{Knoll2010_RDM,Haller1981_PhysicsUPGe,Mertens2019_NIMA_LowNetImpurity}. If an internal p--n junction is present, depletion can initiate simultaneously from multiple locations~\cite{Knoll2010_RDM,Abt2021_JINST_SSD}. In the GeRC field configuration, this can generate competing depletion fronts and leave an isolated region of undepleted material that cannot be eliminated within a practical bias range. To ensure a single monotonic depletion progression from the large-area HV electrode toward the recessed ring contact, we selected an n-type HPGe crystal for the prototypes.

To improve handling yield during aggressive chemical etching and to avoid unintended contact with critical electrode surfaces, shallow wing-and-groove handling features were added near the top and bottom of the detector, as shown in Fig.~\ref{fig:gerc_cross_section}. These features provide mechanically robust clamp points for tweezers and enable controlled flipping during wet processing. They reduce the probability of scratching or contaminating the functional electrode regions. Although they increase machining and polishing complexity, this trade-off is acceptable at the present stage because the immediate objective is to establish a repeatable fabrication workflow for the GeRC topology.

\subsection{Electrostatic modeling with \texttt{SolidStateDetectors.jl}}
Electrostatic simulations were performed using \texttt{SolidStateDetectors.jl} (\texttt{SSD.jl})~\cite{Abt2021_JINST_SSD,Knoll2010_RDM,Haller1981_PhysicsUPGe}, which solves Poisson's equation with a space-charge term determined by the net impurity concentration and applies the prescribed electrode boundary conditions of the detector geometry. The modeled crystal was assigned a uniform net impurity concentration of $N_{\mathrm{eff}}=-2.6\times10^{10}~\mathrm{cm^{-3}}$, where the negative sign denotes n-type bulk. The simulation temperature was set to 78~K to match the liquid-nitrogen operating conditions used for early cryogenic characterization. The ring electrode and the large-area electrode were assigned fixed potentials, and the applied bias was scanned to determine the onset of full depletion.

\begin{figure}[t]
    \centering
    \includegraphics[width=\linewidth]{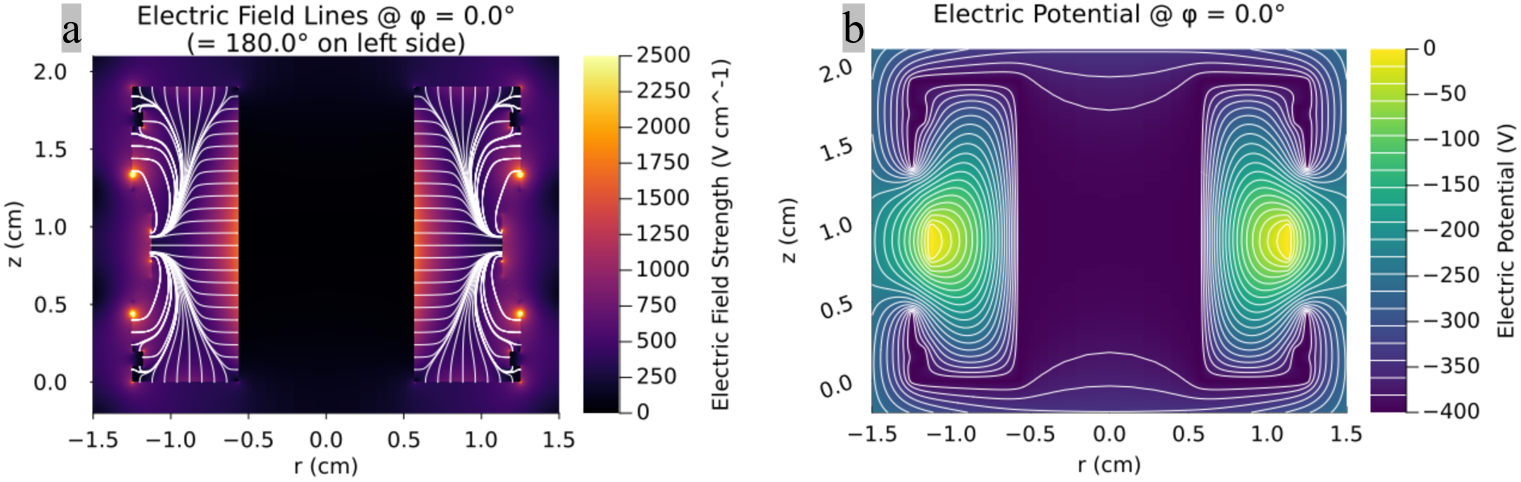}
    \caption{Electrostatic simulation results for the compact GeRC geometry in the $\varphi=0^\circ$ plane, generated using \texttt{SSD.jl}. (a) Electric-field magnitude map with superimposed field lines. The field is concentrated near the recessed ring-contact region and at sharp geometric edges, reflecting the strong local field shaping produced by the small readout electrode and the nonplanar electrode boundaries. (b) Electric-potential map with equipotential contours for the same cross section. The potential distribution shows a smooth bulk gradient from the large-area HV electrode toward the recessed ring contact, while the contour crowding near the groove indicates regions of enhanced field strength. Together, these results illustrate the characteristic GeRC field topology used to guide depletion-voltage estimation and prototype optimization.}
    \label{fig:gerc_field_potential}
\end{figure}

Representative electric-field and electric-potential distributions in the $\varphi=0^\circ$ plane are shown in Fig.~\ref{fig:gerc_field_potential}. Figure~\ref{fig:gerc_field_potential}a presents the electric-field magnitude together with field lines, while Fig.~\ref{fig:gerc_field_potential}b shows the corresponding electric-potential map with equipotential contours. As expected for a small readout electrode recessed into an annular groove, the field is strongly compressed near the ring-contact region and near geometric edges where the electrode boundaries bend sharply. Away from these localized regions, the field extends across the bulk from the large-area HV electrode toward the recessed ring contact, establishing the drift configuration required for charge collection.

For the stated impurity level and the present electrode spacing, the simulations predict a full-depletion voltage in the range of 300--350~V. In this geometry, the depletion voltage is especially sensitive to the ring-contact width and to the separation between the ring electrode and the surrounding HV electrode regions. These electrostatic results therefore served as an important design input for the prototype dimensions and provide a baseline expectation for subsequent cryogenic biasing tests.

\section{GeRC Fabrication Workflow and Process Development}

\subsection{Precision Machining of the Core-and-Groove Geometry: Process Development and Optimization}

All GeRC blanks in this study were produced from an n-type HPGe crystal to ensure a single carrier polarity in subsequent depletion operation. Because detector-grade Ge is mechanically brittle, the machining workflow was optimized to minimize thermal load, torque transients, and localized contact pressure~\cite{Wei2019_JINST_aGeDetectors,Luke2000_OrthStrip,Amman2020_SegmentedAS}. The process targets two outcomes: (i) reproducible core-and-groove geometry, and (ii) damage states that can be fully removed by the downstream polishing and chemical etching sequence~\cite{Wei2019_JINST_aGeDetectors,Luke2000_OrthStrip,Amman2020_SegmentedAS}.

\begin{figure}[t]
    \centering
    \includegraphics[width=\linewidth]{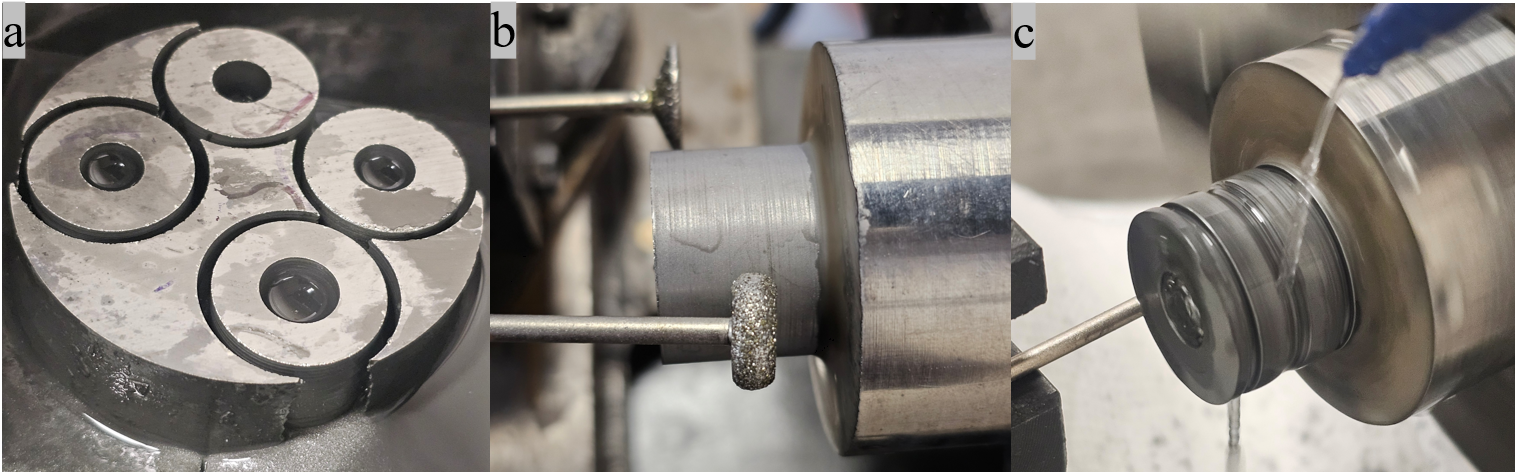}
    \caption{Representative photographs of the early mechanical fabrication sequence for compact GeRC prototypes. (a) Four cylindrical HPGe blanks after cold core drilling and through-bore formation, still supported in the aluminum drilling bath used for coolant circulation and mechanical stabilization. (b) Lathe-based machining of the outer recessed region using a diamond abrasive tool, with the crystal mounted concentrically on a mandrel to maintain alignment with the bore axis. (c) Continued groove and wing machining under direct liquid cooling at the tool--crystal interface. These steps illustrate the low-speed, low-force, coolant-assisted machining strategy adopted to minimize chipping, suppress subsurface crack formation, and preserve the structural integrity of the brittle HPGe crystals during fabrication~\cite{Wei2019_JINST_aGeDetectors,Dong2026_arXiv_HybridProcess,Panamaldeniya2026_arXiv_ICPCaGe}.}
    \label{fig:drill_cut}
\end{figure}

\paragraph{Cold core drilling of cylindrical blanks and through-bore formation.}
A Ge slab was wax-mounted inside a dedicated aluminum bath that served as a coolant reservoir during drilling. Cooling water was continuously circulated through the bath using two small fountain pumps, providing steady flushing of swarf and stabilizing the surface temperature. The bath was rigidly clamped to a drill press, and core drill bits were used to extract four cylindrical blanks with a diameter of 27~mm. Each blank was then cold-drilled to form an 11~mm diameter through-bore. For both steps, low spindle speed and low axial feed force were enforced. This operating window reduced the probability of surface chipping and suppressed the formation of subsurface microcracks that can later seed catastrophic fracture. Representative views of the drilling and subsequent machining workflow are shown in Fig.~\ref{fig:drill_cut}. The drilled cylindrical blanks with through-bores are shown in Fig.~\ref{fig:drill_cut}a.

After drilling, the wax was removed by gentle heating, and each blank was rinsed to remove residual wax and particulate debris. The as-machined surfaces were visually inspected under strong illumination to screen for radial cracks, edge spallation, or bore-wall delamination before proceeding.

\paragraph{Lathe-based groove and wing machining with low-risk tooling.}
To machine the recessed ring region and the two wing regions, each blank was mounted concentrically on a stainless-steel mandrel. The small Ge core recovered from the bore-drilling step was reinserted into the through-bore and temporarily fixed in place to distribute cutting loads and reduce stress concentration on the bore wall. The mandrel assembly was held in a lathe chuck, and the crystal was centered using a dial test indicator to ensure rotation about the bore axis. Figures~\ref{fig:drill_cut}b and \ref{fig:drill_cut}c show representative views of this machining stage, including the abrasive cutting tool, the rotating mandrel-mounted crystal, and the application of liquid cooling directly at the tool--Ge interface.

Multiple tool geometries and abrasive materials were evaluated during early trials. Based on observed failure modes, diamond grinding bits were adopted as the primary machining tool because they provided the most stable material removal with the lowest incidence of edge breakout. Localized pits and surface pull-out were most likely to appear after the tool had advanced to depth or during tool retraction, consistent with friction-driven damage in a hard but brittle covalent crystal. Two mitigation measures were implemented throughout this stage: continuous water spray at the tool--Ge interface, and frequent light abrasion of the freshly cut surface using fine sandpaper. This combination reduced tool loading, improved surface finish during cutting, and lowered the probability of spallation at geometric transitions.

Following completion of the groove and wing features, the blanks were removed from the mandrel and re-inspected for cracks and chipping.

\subsection{Surface Conditioning: core-and-groove Mechanical Polishing and Chemical Damage Removal}

\begin{figure}[t]
    \centering
    \includegraphics[width=0.78\linewidth]{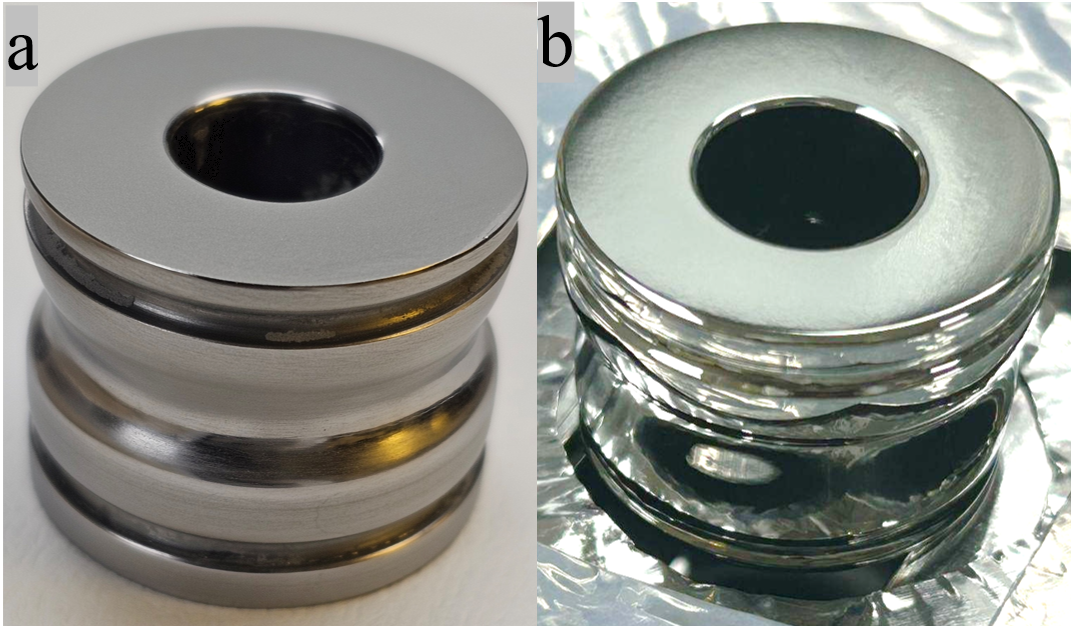}
    \caption{Photographs of a compact GeRC prototype after the two final stages of surface preparation. (a) Detector appearance immediately after completion of the full mechanical polishing sequence, including edge rounding, bore polishing, curved-surface sanding, and planar-face lapping. At this stage, the geometry is fully defined and the surface appears smooth and reflective, but a mechanically damaged near-surface layer may still remain. (b) Detector appearance after the subsequent heavy chemical etch. This step removes the residual mechanically damaged layer and produces a more chemically uniform surface, visible here as a less reflective and more matte finish. The post-etch condition shown in panel (b) was used as the prepared surface state for subsequent detector-fabrication steps.}
    \label{fig:polish_etch}
\end{figure}

Surface preparation for the GeRC topology must address both visible machining defects and the subsurface damage layer introduced during drilling, grinding, and groove machining, since both can increase surface leakage and degrade high-voltage stability. The workflow therefore combined staged mechanical polishing with a controlled chemical polish etch~\cite{Morgan1969_GeLiPlanar,Wei2019_JINST_aGeDetectors,Panth2020_EPJC_Cryo}. Representative surface appearances after the final mechanical polishing step and after the subsequent heavy etch are shown in Fig.~\ref{fig:polish_etch}. Figure~\ref{fig:polish_etch}a shows the detector immediately after completion of mechanical polishing, when the surfaces were already visually smooth and geometrically well defined. Figure~\ref{fig:polish_etch}b shows the same detector after the heavy etch, which removed the mechanically damaged surface layer and produced the more chemically equilibrated matte appearance used as the starting point for later contact processing.

\paragraph{Coarse defect removal.}
After groove and wing machining, localized pits and deep scratches were first removed using a 60-grit grinding bit. This step was restricted to visibly damaged regions and geometric transition edges, where machining defects were most likely to persist. Material removal was kept to the minimum required to eliminate the defect features while preserving the target dimensions of the crystal.

\paragraph{Progressive sanding of edges, bore, and curved sidewalls.}
A grit sequence from 120 to 2500 was then used to reduce surface roughness in a controlled manner. Sharp edges at the top and bottom faces were rounded early in the sequence to suppress edge chipping during later handling and during the chemical etch~\cite{Morgan1969_GeLiPlanar,Wei2019_JINST_aGeDetectors,Jany2021_EPJC_Prototype}. The central bore and recessed groove regions were polished using sanding drill bits with the same grit progression. The outer cylindrical sidewall was sanded with progressively finer grit while maintaining a consistent polishing direction. This reduced cross-scratching and helped avoid nonuniform local pressure that could imprint new grooves on the curved surfaces.

\paragraph{Planar-face lapping.}
The top and bottom faces were finished on a flat glass plate. Coarse lapping used 20~$\mu$m lapping powder to remove residual sanding marks and recover planar surfaces. Fine lapping then used 10~$\mu$m lapping powder to suppress the remaining fine scratches. A figure-eight motion was used throughout to average hand pressure and limit preferential wear~\cite{Wei2019_JINST_aGeDetectors,Panth2022_NIMA_TempBarrier,Amman2020_SegmentedAS}. Lapping continued until no scratches were visible under inspection from multiple lighting angles.

\paragraph{Chemical polish etch and iterative inspection.}
Chemical damage removal was performed using a HNO$_3$:HF mixture at a 4:1 volumetric ratio. Each etch cycle lasted 3~min, followed immediately by a DI-water rinse, drying with nitrogen gas, and visual inspection~\cite{Wei2019_JINST_aGeDetectors,Panth2022_NIMA_TempBarrier,Dong2026_arXiv_HybridProcess}. If isolated scratches or pits remained, the affected region was locally returned to the appropriate sandpaper grit and then re-etched. This polish--etch--inspect cycle was repeated until all surfaces appeared uniformly smooth, or exhibited a uniform orange-peel texture, with no pits or scratches visible to the unaided eye. The visual contrast between Fig.~\ref{fig:polish_etch}a and Fig.~\ref{fig:polish_etch}b reflects this transition from the mechanically polished state to the post-heavy-etch surface condition.

\subsection{Conformal a-Ge Thin-Film Encapsulation via RF Sputter Deposition}

\begin{figure}[tbp]
    \centering
    \includegraphics[width=\textwidth]{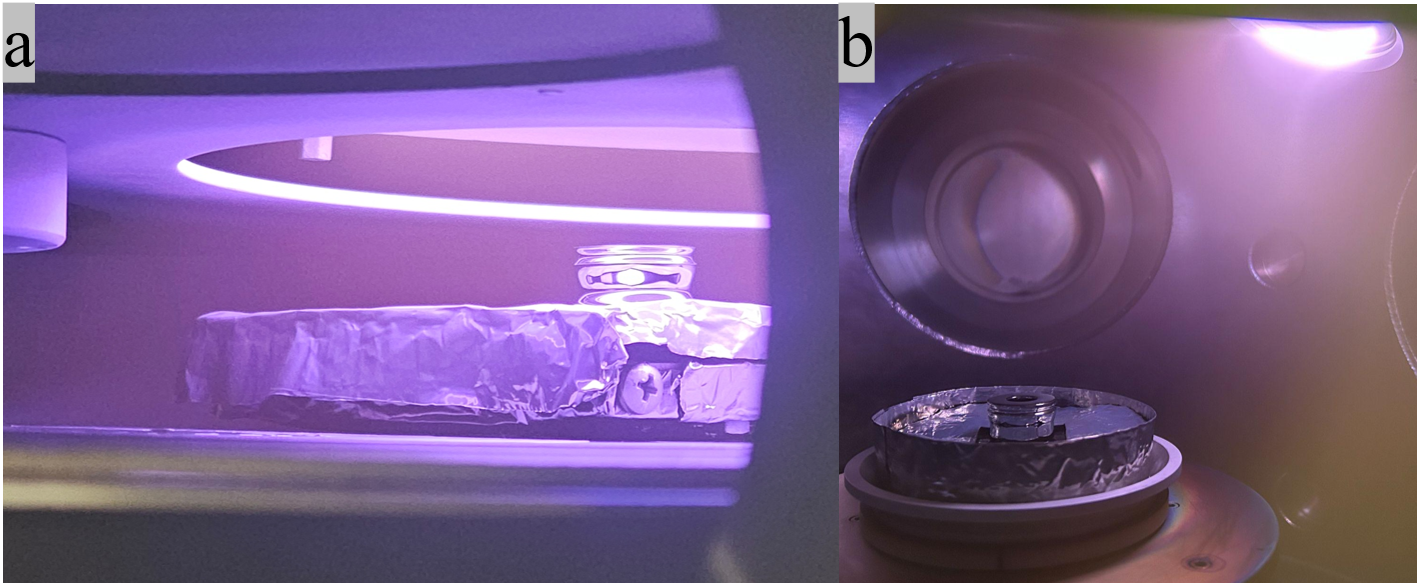}
    \caption{Two sputter-deposition systems used for fabrication of compact HPGe detector prototypes in this work. Panel (a) shows the older Perkin--Elmer sputtering system, and panel (b) shows the AJA RF/DC magnetron sputtering system. The two platforms differ in several hardware aspects that are relevant to film growth on detector surfaces, including the target--substrate distance, target size, and vacuum pumping configuration. The Perkin--Elmer system provides a shorter deposition throw and larger, 8-inch-class targets, whereas the AJA system uses 3-inch-class targets in a different chamber geometry. The systems also differ in high-vacuum pumping, with a helium cryopump in the Perkin--Elmer tool and a turbomolecular pump in the AJA tool. Comparing detectors fabricated on these two systems provides a practical test of whether the a-Ge deposition and masking procedure remains operable across substantially different sputtering environments. The comparison is intended as a robustness check for the workflow rather than as a full demonstration of process equivalence between the two tools.}
    \label{fig:sputtering_systems}
\end{figure}

To test the robustness of the contact-fabrication workflow against hardware-dependent deposition conditions, two compact HPGe GeRC-related devices were processed on two independent sputter-deposition platforms. One detector was fabricated using an older Perkin--Elmer sputtering system, while the other was processed using an AJA RF/DC magnetron sputtering system. As shown in Fig.~\ref{fig:sputtering_systems}, the two tools differ substantially in chamber geometry and vacuum architecture. The most important hardware differences are the target--substrate spacing, the target size, and the high-vacuum pumping configuration. In particular, the Perkin--Elmer system uses a shorter target throw distance and larger, 8-inch-class targets, together with a helium cryopump, whereas the AJA system employs 3-inch-class targets and a turbomolecular pumping configuration. These differences primarily affect the deposition geometry and the practical strategy required to obtain sufficiently uniform and conformal amorphous-Ge coverage on recessed or non-planar detector features~\cite{Thornton1974_JVST_SputterGeometry,Hull2005_aGeContacts,Amman2020_SegmentedAS,Amman2018_aGeOptimization}.

Immediately before a-Ge deposition, each detector received a brief pre-deposition etch (30~s) to remove surface oxide/contamination, disturbed near-surface material, and surface contamination left by the preceding machining and polishing steps~\cite{Wei2019_JINST_aGeDetectors,Amman2020_SegmentedAS,Dong2026_arXiv_HybridProcess}. After the etch, the crystal was rinsed thoroughly in DI water and dried using high-purity nitrogen. To minimize the risk of introducing new defects during post-etch handling, the detector body was not touched directly by any tools. A dedicated tweezer was used only at the wing or handle region, thereby avoiding mechanical contact with active surfaces and preserving the freshly prepared sidewalls, top face, and recessed geometrical features.

The detector was then mounted on a dedicated fixture for sputter processing. Aluminum-foil masking was applied around the jig to shield areas that were not intended to receive film deposition and to suppress line-of-sight coating of unintended surfaces. This masking step was particularly important for maintaining control of the final contact geometry in regions adjacent to the ring structure and other non-planar features, where uncontrolled overspray could complicate later pattern definition or alter the effective electrical boundary conditions~\cite{Thornton1974_JVST_SputterGeometry,Amman2020_SegmentedAS,Amman2018_aGeOptimization}.

A conformal a-Ge encapsulation layer was deposited by RF sputtering from a Ge target in an Ar/H$_2$ gas mixture ~\cite{Hull2005_aGeContacts,Amman2020_SegmentedAS,Amman2018_aGeOptimization}. To improve film coverage on recessed, partially shadowed, or geometrically obstructed surfaces, the deposition was carried out sequentially from two opposing directions. After the first deposition, the detector was removed from the chamber, flipped while being handled only at the wing region, remounted on the fixture, and sputtered again from the opposite orientation. Before each deposition step, a pre-sputter was performed with the substrate shutter closed in order to remove surface oxide from the target, stabilize the plasma, and establish a steady deposition rate before the detector surface was exposed~\cite{Utrera2018_TSF_TargetCleaning,GarciaGancedo2012_JAP_PreDeposition,Kossoy2015_JVSTA_ShutterStability}.

The RF sputter parameters and timing sequences for both systems are summarized in Table~\ref{tab:age_sputter_params}. For the Perkin--Elmer system, deposition was segmented into multiple short runs with intermediate cool-down intervals to limit substrate heating during the short-throw process. For the AJA system, a single continuous deposition was used. Because the AJA system uses a smaller target, the substrate height and the magnetron gun angle were adjusted to improve thickness uniformity across the detector footprint~\cite{Thornton1974_JVST_SputterGeometry,Amman2020_SegmentedAS,Amman2018_aGeOptimization}.

\begin{table}[t]
  \centering
  \caption{RF sputter parameters for conformal a-Ge deposition in the two sputtering systems. The Ar/H$_2$ mixture contained 7\% H$_2$ in Ar. Base pressure refers to the overnight pump-down prior to initiating the process gas flow.}
  \label{tab:age_sputter_params}
  \begin{tabular}{lcc}
    \hline
    Parameter & Perkin--Elmer system & AJA system \\
    \hline
    Base pressure (overnight) & $3\times 10^{-6}$ Torr & $3\times 10^{-7}$ Torr \\
    RF power & 100 W & 200 W \\
    Working pressure & 14 mTorr & 3 mTorr \\
    Process gas flow & 65 sccm (Ar/7\% H$_2$) & 20 sccm (Ar/7\% H$_2$) \\
    Pre-sputter (shutter closed) & 5 min (initial), 2 min (each cycle) & 10 min \\
    Deposition time & 5$\times$3 min (10 min cool-down) & 15 min (continuous) \\
    Deposition directions & two-sided (flip and repeat) & two-sided (flip and repeat) \\
    \hline
  \end{tabular}
\end{table}

\subsection{Al Metallization for Ohmic Electrode Formation by Sputtering}

Following the a-Ge encapsulation, aluminum metallization was deposited to define the electrode areas and to provide a low-resistance metallic layer suitable for electrical connection~\cite{Panth2020_EPJC_Cryo,Wei2019_JINST_aGeDetectors,Yang2022_WraparoundLiAGe}. The Al layer was deposited by DC magnetron sputtering. The handling protocol and fixturing strategy mirrored the a-Ge step: the detector was transferred using tweezers contacting only the wing/handle region, mounted on the jig, and masked with aluminum foil as needed to protect non-electrode surfaces. A pre-sputter step with the shutter closed was performed to clean the Al target and stabilize the discharge before opening the shutter for deposition~\cite{Schelfhout2015_ApplSurfSci_TargetCleanness,Utrera2018_TSF_TargetCleaning,Kossoy2015_JVSTA_ShutterStability}.

The DC sputter parameters for Al deposition are summarized in Table~\ref{tab:al_sputter_params}. In both systems, ultra-high-purity Ar was used as the sputter gas and the working pressure was held at 3~mTorr. Consistent with the smaller target size in the AJA system, the substrate height and gun angle were adjusted during the AJA deposition to improve coating uniformity across the detector surface.

\begin{table}[t]
  \centering
  \caption{DC sputter parameters for Al metallization in the two sputtering systems. Base pressure refers to the overnight pump-down prior to initiating the Ar flow.}
  \label{tab:al_sputter_params}
  \begin{tabular}{lcc}
    \hline
    Parameter & Perkin--Elmer system & AJA system \\
    \hline
    Base pressure (overnight) & $3\times 10^{-6}$ Torr & $3\times 10^{-7}$ Torr \\
    DC power & 300 W & 400 W \\
    Working pressure & 3 mTorr & 3 mTorr \\
    Process gas flow & 19.5 sccm (UHP Ar) & 20 sccm (UHP Ar) \\
    Pre-sputter (shutter closed) & 6 min & 10 min \\
    Deposition time & 6 min & 8 min \\
    \hline
  \end{tabular}
\end{table}

\subsection{Ring-Contact Patterning: Electrode Layout Design and Definition by Light Chemical Etching}

After sputter metallization, the Al layer initially covers all a-Ge-encapsulated surfaces and must be patterned to establish the final electrode geometry~\cite{Luke1992_aGeBipolar,Hull2005_aGeContacts,Wei2019_JINST_aGeDetectors,Luke2000_OrthStrip,Amman2020_SegmentedAS,Yang2022_WraparoundLiAGe}. In the GeRC topology, this step serves two purposes: it electrically isolates the narrow ring readout electrode from the large-area high-voltage electrode, and it re-exposes the intended a-Ge passivation bands on the groove sidewall and adjacent surfaces. Both GeRC prototypes were patterned using the same masking-and-light-etch procedure, with the principal design difference being the retained ring width. The device processed in the older Perkin--Elmer sputtering system retained a ring contact of approximately $0.8$~mm, whereas the device processed in the AJA system retained a wider ring contact of approximately $2$~mm.

\begin{figure}[tbp]
    \centering
    \includegraphics[width=\textwidth]{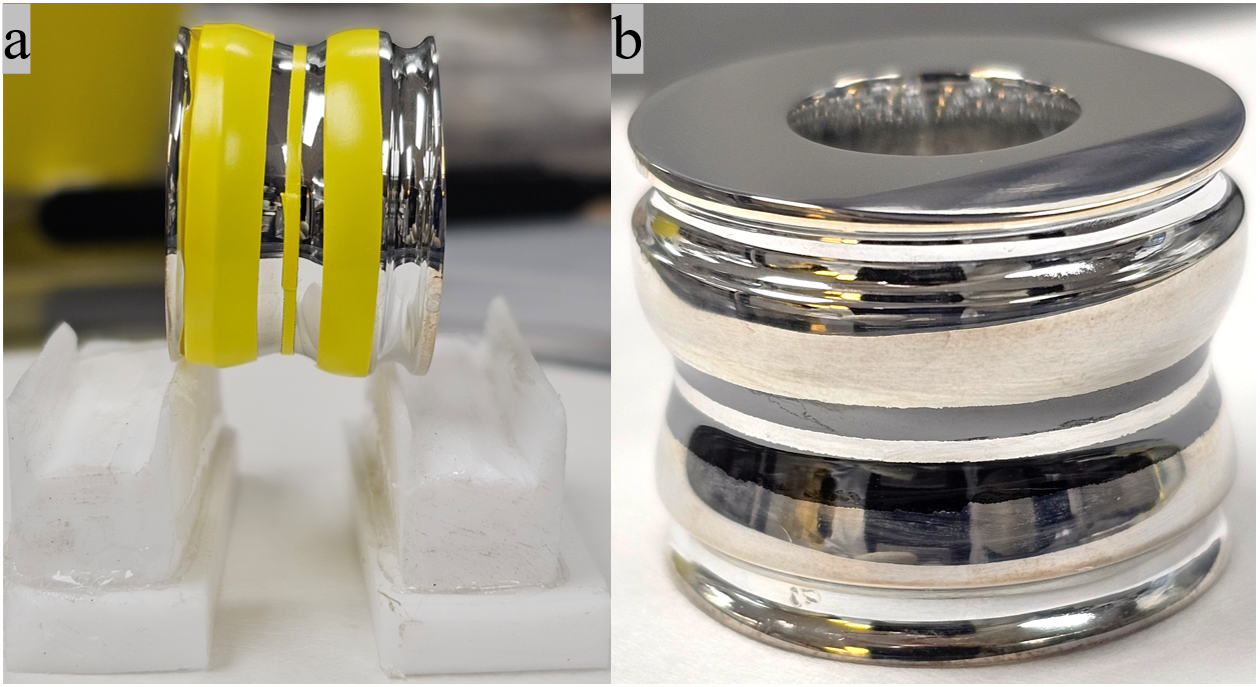}
    \caption{Masking and final appearance of a compact GeRC prototype during Al patterning after sputter metallization. Panel (a) shows the detector mounted on a PTFE rail holder with 3M~471 vinyl tape applied to define the electrode boundaries and protect the metallized regions that were intended to remain after etching. This fixture provided mechanical support and stand-off spacing during the light HF etch, allowing liquid to drain from the non-planar detector surfaces rather than accumulate in the groove region. Panel (b) shows the detector after completion of the Al patterning step and removal of the masking tape. The final structure exhibits electrically separated metallized regions and exposed intermediate bands corresponding to the intended a-Ge passivation and isolation regions. This patterning step defines the ring-contact readout geometry and removes unwanted Al bridges that would otherwise short the narrow ring electrode to the large-area high-voltage contact.}
    \label{fig:light_etch_masking}
\end{figure}

The electrode boundaries were defined directly on the core-and-groove geometry using a mechanical mask made from 3M~471 vinyl tape. Narrow tape strips were first cut and applied around the groove region to define the two edges of the ring electrode. The tape placement was aligned to the groove centerline, and the resulting ring width was verified with a caliper before wet etching. Additional tape pieces were then applied to protect the electrode regions intended to remain metallized, while leaving the Al in the isolation and passivation bands exposed for removal. As shown in Fig.~\ref{fig:light_etch_masking}, the masked detector was supported on a simple PTFE rail holder during this step. Because the groove and wing features introduce curved surfaces, local slopes, and short vertical sections, careful burnishing of the tape edges was required to achieve intimate contact and to suppress undercutting or etchant migration beneath the mask.

Selective Al removal was carried out using a light HF etch prepared from a dilute 100:1 HF solution, corresponding to approximately 0.49\% $\mathrm{HF}$ in water~\cite{Panth2022_NIMA_TempBarrier,Panth2020_EPJC_Cryo,Amman2020_SegmentedAS}. During the etch, the masked detector was placed on the PTFE holder shown in Fig.~\ref{fig:light_etch_masking}(a), which acted as a sliding rail and stand-off support. This configuration allowed the detector to be lowered and manipulated such that the etchant drained rather than pooled on the groove and adjacent non-planar surfaces. HF was applied with a disposable dropper directly onto the exposed Al regions, and the detector was rotated slowly to maintain continuous wetting around the full groove circumference. The etch was stopped as soon as the exposed Al lost its metallic luster and the underlying a-Ge surface became visually uniform.

After etching, the detector was rinsed thoroughly with DI water and dried with high-purity nitrogen. The vinyl tape was then removed carefully to avoid lifting, smearing, or scratching the thin-film layers. The patterned detector was subsequently inspected for residual Al bridges or narrow metallic filaments spanning the isolation bands. When such remnants were observed, brief local touch-up etches were applied to restore complete electrical isolation between the ring electrode and the high-voltage electrode. The final patterned appearance is shown in Fig.~\ref{fig:light_etch_masking}(b).

\section{Cryogenic Characterization and Performance Evaluation of GeRC Prototypes}

\subsection{Cryogenic Measurement Platform, GeRC-Specific Mount, and Readout Configuration}

The compact GeRC prototypes were characterized at liquid-nitrogen temperature using the same vacuum cryogenic test platform previously employed for planar-device studies~\cite{Dong2026_arXiv_HybridProcess,Panth2020_EPJC_Cryo,Jany2021_EPJC_Prototype}. The cryostat provides a high-vacuum environment that suppresses convective heat load and surface condensation, enabling stable detector operation near $77~\mathrm{K}$ during electrical and spectroscopic measurements~\cite{Knoll2010_RDM,Haller1981_PhysicsUPGe,Bonet2021_EPJC_CONUS}.

\begin{figure}[tbp]
    \centering
    \includegraphics[width=\textwidth]{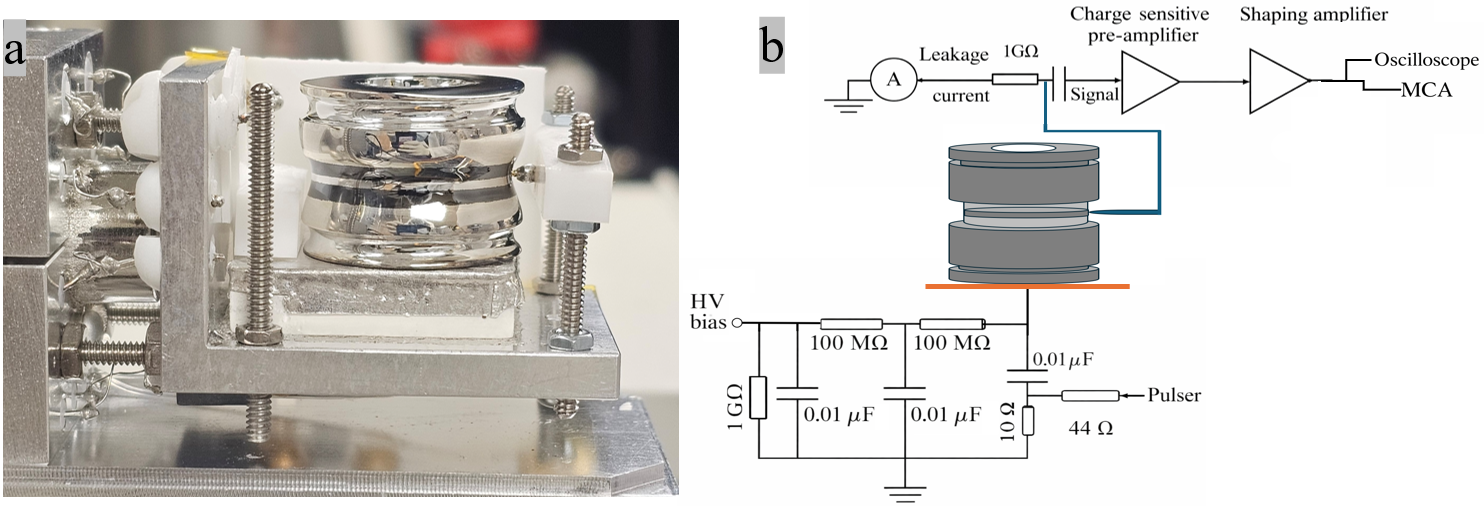}
    \caption{GeRC-specific cryogenic test configuration and readout chain used for electrical and spectroscopic characterization. Panel (a) shows a compact GeRC prototype mounted in the modified cryostat fixture. The detector is supported on a PTFE-based platform, while a laterally approaching pogo-pin contact is used to read out the recessed ring electrode located inside the annular groove. A second PTFE support piece on the opposite side provides controlled mechanical preload and helps maintain stable electrical contact during cooldown and operation at liquid-nitrogen temperature. Panel (b) shows the corresponding measurement schematic. The large-area metallized electrode network was used as the high-voltage contact, biased through a resistor and filter chain, while the recessed ring electrode served as the signal readout. The front-end electronics consisted of a charge-sensitive preamplifier followed by a shaping amplifier, with outputs sent to an oscilloscope and a multichannel analyzer (MCA). A calibrated pulser input was included to monitor electronic noise and gain stability during operation, and the leakage current was measured in the bias circuit.}
    \label{fig:gerc_setup}
\end{figure}

Because the GeRC geometry differs substantially from conventional planar or point-contact layouts, a GeRC-specific mechanical and electrical interface was developed for cryogenic testing. In the original configuration of the test platform, the signal electrode was contacted from above using a spring-loaded pogo-pin assembly. That arrangement is not well suited to the present devices, because the GeRC readout electrode is a narrow metallized ring recessed inside the annular groove. A top-down approach is therefore mechanically inconvenient and comparatively sensitive to alignment error. To address this issue, the cryostat was modified to use a lateral pogo-pin contact that approaches the ring electrode from the side, as shown in Fig.~\ref{fig:gerc_setup}(a).

The lateral pogo-pin assembly is mounted on a PTFE support structure that provides electrical insulation at cryogenic temperature while maintaining sufficient mechanical rigidity. Its vertical position is adjustable, allowing the spring tip to be aligned to the ring-contact height and centered within the groove. After coarse alignment, a second PTFE piece is inserted from the opposite side of the detector to apply a controlled preload. This arrangement presses the detector gently but reproducibly against the pogo pin, improving contact stability while avoiding unnecessary shear force or direct handling of the thin-film contact surfaces.

Following installation, the cryostat vessel was evacuated to high vacuum in the $10^{-6}~\mathrm{Torr}$ range. After the base pressure was reached, liquid nitrogen was transferred into the dewar to initiate cooldown. The system was then held under vacuum and cooled overnight so that the detector mount, contact structures, and surrounding thermal mass could reach equilibrium before biasing and data acquisition. The full cryogenic mounting concept and associated front-end readout chain are summarized in Fig.~\ref{fig:gerc_setup}.

\subsection{Electrical and Spectroscopic Characterization of the Two GeRC Prototypes}

With the GeRC-specific cryogenic interface established, the two compact prototypes were then evaluated electrically and spectroscopically at liquid-nitrogen temperature using the readout chain shown in Fig.~\ref{fig:gerc_setup}(b). For both devices, the high-voltage electrode corresponded to the large-area interconnected metallized regions, including the bore, the top and bottom faces, and the connected outer sidewall segments. The recessed ring electrode inside the annular groove was used as the signal readout contact.

The primary purpose of this first-generation study was not yet to optimize ultimate detector performance, but rather to establish whether the compact GeRC topology could be fabricated, mounted, biased, and read out reproducibly under cryogenic conditions. In particular, the measurements were intended to verify that the core-and-groove machining and polishing sequence does not preclude stable low-temperature operation, and that conformal thin-film contacts can function on the non-planar GeRC surface prior to the introduction of lithium-diffused contacts in later device generations.

The measurements described below therefore focus on the fundamental operational questions for these prototypes, including leakage-current behavior, high-voltage stability, and the ability to obtain spectroscopic response with the recessed ring-contact readout geometry. Together, these tests provide the first experimental benchmark for compact GeRC devices fabricated with fully sputtered thin-film contact structures.

\paragraph{Biasing stability, leakage-current behavior, and depletion proxy.}
Leakage current was recorded as a function of the applied bias magnitude after the detector reached thermal equilibrium at liquid-nitrogen temperature~\cite{Jany2021_EPJC_Prototype,Agostini2019_EPJC_BEGeChar,Bonet2021_EPJC_CONUS}. In parallel, the pulser response was tracked as a function of bias and used to construct an effective-capacitance proxy for the detector-plus-front-end input node~\cite{Panth2020_EPJC_Cryo,Jany2021_EPJC_Prototype,Agostini2019_EPJC_BEGeChar}. For a fixed injected charge, the measured pulser amplitude changes with the total input capacitance, so the bias dependence of this quantity provides a practical indicator of depletion evolution even without a separate absolute capacitance calibration~\cite{Panth2020_EPJC_Cryo,Jany2021_EPJC_Prototype,Bonet2021_EPJC_CONUS}.

\begin{figure}[tbp]
    \centering
    \includegraphics[width=\textwidth]{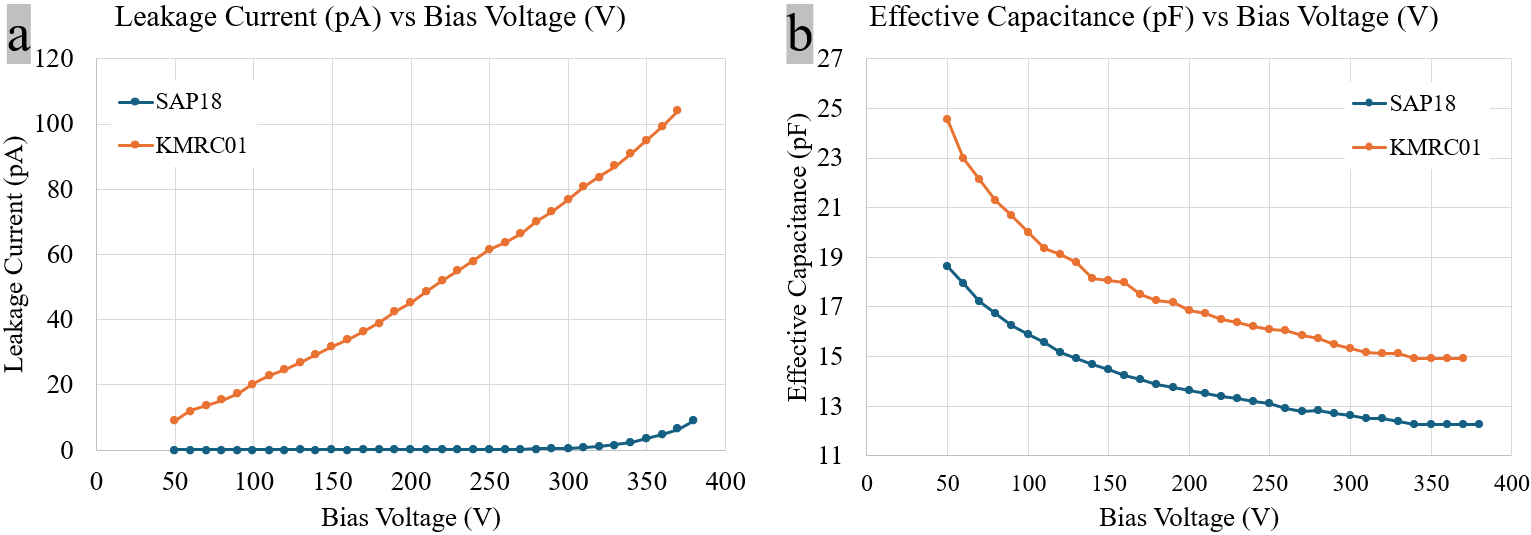}
    \caption{Comparison of the low-temperature electrical characteristics of the two compact GeRC prototypes. Panel (a) shows the measured leakage current as a function of applied bias for SAP18 and KMRC01. SAP18, fabricated using the older Perkin--Elmer sputtering system, maintains very low leakage current over most of the scanned range and shows a noticeable rise only near the highest tested biases. KMRC01, fabricated using the AJA system, exhibits substantially larger leakage current across the full scan and increases approximately monotonically with bias. Panel (b) shows the corresponding effective-capacitance proxy versus bias for the same two detectors, derived from the pulser response of the detector-plus-front-end input node. In both devices, the quantity decreases with increasing bias and becomes progressively flatter at higher voltage, consistent with depletion extending through the detector volume and approaching a near-plateau regime. The comparison shows that both devices are operable as GeRC prototypes, but that SAP18 provides a substantially larger low-leakage operating margin than KMRC01.}
    \label{fig:ivcv_compact_gerc}
\end{figure}

The leakage-current and effective-capacitance behaviors of the two compact GeRC prototypes are compared directly in Fig.~\ref{fig:ivcv_compact_gerc}. The Perkin--Elmer-processed prototype, SAP18, shows very low leakage over most of the scanned voltage range. As shown in Fig.~\ref{fig:ivcv_compact_gerc}(a), the current remains near the baseline level through roughly the first $250$--$300~\mathrm{V}$ of the scan and then increases only near the highest tested biases, reaching about $9~\mathrm{pA}$ at the upper end of the measurement range. The corresponding effective-capacitance proxy in Fig.~\ref{fig:ivcv_compact_gerc}(b) decreases monotonically with bias and becomes essentially constant above $\sim 340~\mathrm{V}$, consistent with the detector approaching full depletion near that voltage. Because this observable is derived from the pulser response of the detector-plus-front-end node rather than from an absolute capacitance calibration, we interpret the flattening near $340~\mathrm{V}$ as the inferred onset of full depletion rather than as a precision determination of the depletion voltage~\cite{Jany2021_EPJC_Prototype,Agostini2019_EPJC_BEGeChar,Panth2020_EPJC_Cryo}. Based on this behavior, an operating voltage of $380~\mathrm{V}$ was adopted for SAP18 in subsequent measurements.

By contrast, the AJA-processed prototype, KMRC01, exhibits substantially higher leakage current over the entire bias range. The current is already about $9~\mathrm{pA}$ at $50~\mathrm{V}$ and rises steadily with increasing bias, reaching about $104~\mathrm{pA}$ near $370~\mathrm{V}$, as shown in Fig.~\ref{fig:ivcv_compact_gerc}(a). Unlike SAP18, KMRC01 showed small, near-circular grain-boundary-like features on both the top and bottom surfaces after processing. These localized defects are the most likely origin of the elevated leakage, because they can introduce enhanced surface conduction pathways and defect-assisted carrier injection~\cite{Hull2005_aGeContacts,Wei2020_EPJC_CBH,Panth2022_NIMA_TempBarrier}. The corresponding effective-capacitance proxy decreases rapidly at low bias and then gradually flattens, becoming essentially constant above $\sim 340~\mathrm{V}$, consistent with KMRC01 also approaching full depletion near that voltage despite its much higher leakage current~\cite{Jany2021_EPJC_Prototype,Agostini2019_EPJC_BEGeChar,Panth2020_EPJC_Cryo}. As for SAP18, this interpretation is based on the plateau behavior of a pulser-derived proxy rather than on a direct capacitance measurement. Accordingly, an operating voltage of $370~\mathrm{V}$ was selected for KMRC01.

The difference between the high-bias effective-capacitance values of the two detectors should not be interpreted as evidence for fundamentally different depletion behavior. Instead, it mainly reflects geometric differences between the prototypes, specifically small variations in detector volume and the different retained ring-contact widths. Both factors modify the electrode geometry and weighting-field configuration, and therefore lead to different asymptotic effective-capacitance values after the detectors approach full depletion~\cite{Luke1989_IEEE_ShapedField,Cooper2012_NIMA_HPGeTrackingImaging,Abt2021_JINST_SSD}.

\paragraph{Gamma-ray spectroscopy and resolution metrics.}
Representative cryogenic spectra were acquired with $^{241}\mathrm{Am}$ and $^{137}\mathrm{Cs}$ at the operating biases established from the electrical characterization, namely $380~\mathrm{V}$ for SAP18 and $370~\mathrm{V}$ for KMRC01. In each run, a pulser peak was recorded simultaneously under the same shaping and gain conditions so that the electronics contribution to the measured peak width could be estimated directly~\cite{Knoll2010_RDM,Jany2021_EPJC_Prototype,Bonet2021_EPJC_CONUS}. Peak positions and FWHM values reported in the figures and in Table~\ref{tab:gerc_spec_metrics} were obtained from fitted full-energy and pulser peaks within the corresponding acquisition window. Because the low-energy and high-energy source measurements were acquired with different gain windows, the pulser peak appears at different calibrated energies in different runs and should be interpreted only as a reference for the electronic response under the corresponding acquisition setting.

\begin{figure}[t]
  \centering
  \includegraphics[width=0.96\textwidth]{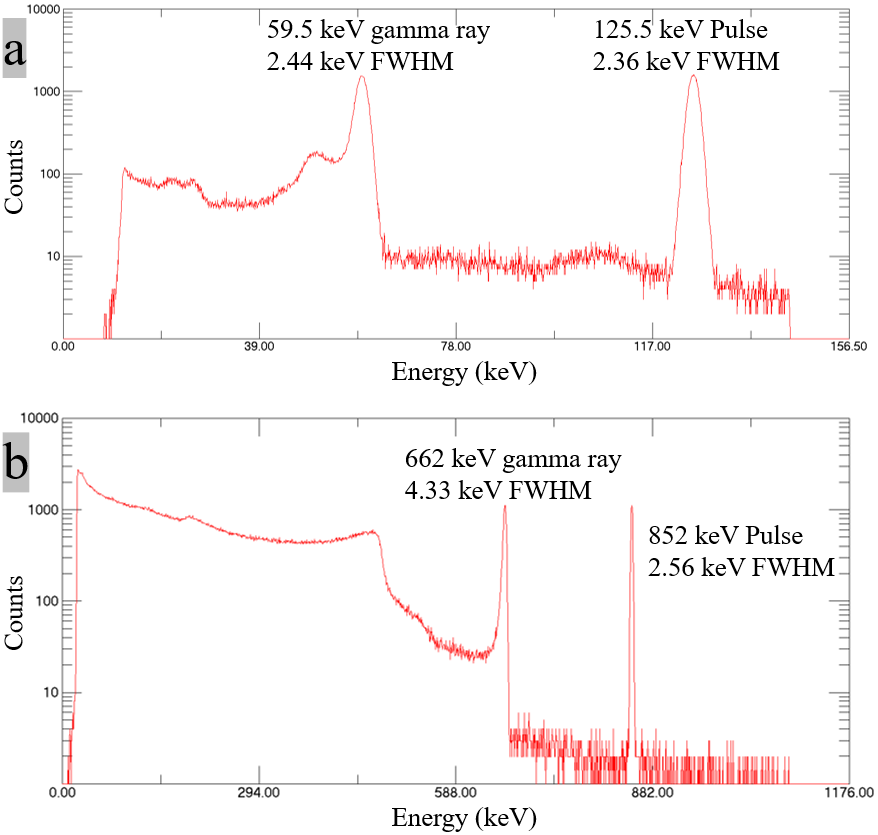}
  \caption{Representative gamma-ray spectra obtained with the SAP18 compact GeRC prototype at $77~\mathrm{K}$, measured at an operating bias of $380~\mathrm{V}$ with a shaping time of $0.5~\mu\mathrm{s}$. Panel (a) shows the $^{241}\mathrm{Am}$ spectrum, including the $59.5~\mathrm{keV}$ full-energy peak and the simultaneously recorded pulser peak displayed at a calibrated energy of $125.5~\mathrm{keV}$ under that acquisition setting. Panel (b) shows the $^{137}\mathrm{Cs}$ spectrum, including the $662~\mathrm{keV}$ full-energy peak and the simultaneously recorded pulser peak displayed at $852~\mathrm{keV}$. The ordinate is plotted on a logarithmic scale to show the full-energy peaks, continuum structure, and baseline region within a single view. The FWHM values labeled in the figure are the fitted quantities used for the resolution analysis summarized in Table~\ref{tab:gerc_spec_metrics}.}
  \label{fig:spec_sap18}
\end{figure}

SAP18 spectra are shown in Fig.~\ref{fig:spec_sap18}. At $380~\mathrm{V}$ with a shaping time of $0.5~\mu\mathrm{s}$, the $59.5~\mathrm{keV}$ full-energy peak from $^{241}\mathrm{Am}$ exhibited a measured FWHM of $2.44~\mathrm{keV}$, while the simultaneously recorded pulser peak showed $2.36~\mathrm{keV}$ FWHM, as seen in Fig.~\ref{fig:spec_sap18}(a). The near equality of these two widths indicates that, under this shaping condition, the observed low-energy resolution is dominated by the front-end electronics~\cite{Knoll2010_RDM,Jany2021_EPJC_Prototype,Bonet2021_EPJC_CONUS}. For the $^{137}\mathrm{Cs}$ run in Fig.~\ref{fig:spec_sap18}(b), the $662~\mathrm{keV}$ photopeak broadened to $4.33~\mathrm{keV}$ FWHM, compared with a pulser width of $2.56~\mathrm{keV}$. After subtraction of the pulser term in quadrature, the remaining detector-related broadening term is $3.49~\mathrm{keV}$, substantially larger than the corresponding value of $0.62~\mathrm{keV}$ at $59.5~\mathrm{keV}$~\cite{Knoll2010_RDM,Jany2021_EPJC_Prototype,Bonet2021_EPJC_CONUS}. This behavior indicates that SAP18 is affected by a significant non-electronic broadening contribution at high energy under the present short-shaping setting. In the present compact GeRC geometry, plausible contributors include electric-field nonuniformity near the groove and bore region, depth-dependent charge-collection times, and local non-idealities in thin-film coverage on recessed or curved surfaces~\cite{Comellato2021_EPJC_Collective,Agostini2021_EPJC_ICPC,Mei2020_JPG_ChargeTrapping}.

\begin{figure}[t]
  \centering
  \includegraphics[width=0.96\textwidth]{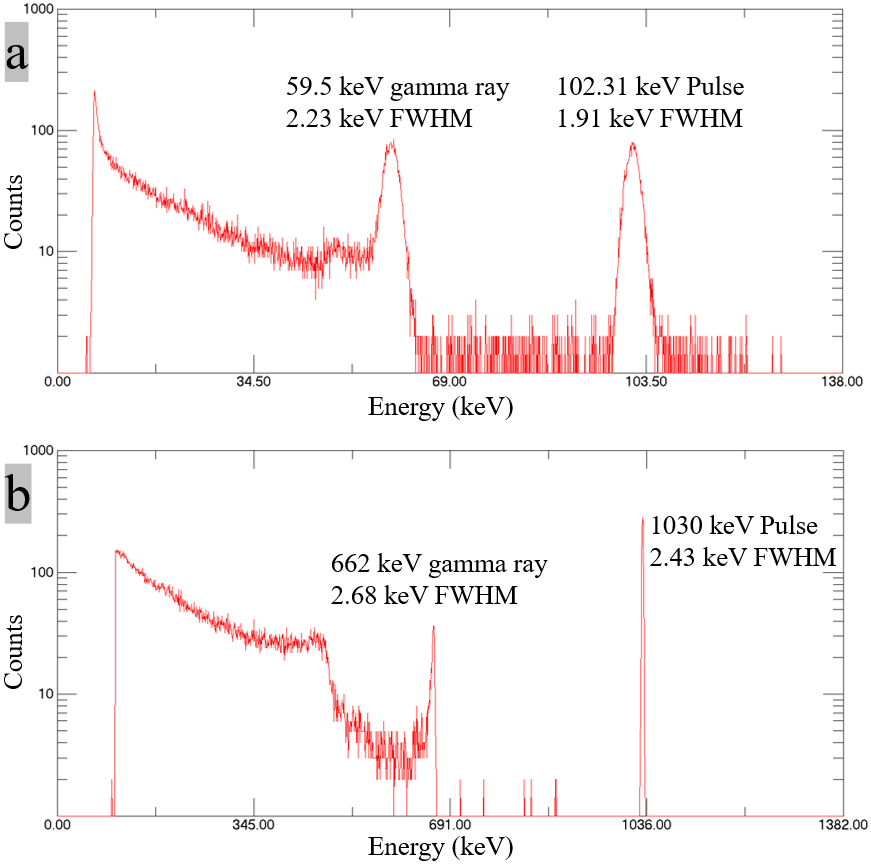}
  \caption{Representative gamma-ray spectra obtained with the KMRC01 compact GeRC prototype at $77~\mathrm{K}$, measured at an operating bias of $370~\mathrm{V}$ with a shaping time of $0.5~\mu\mathrm{s}$. Panel (a) shows the $^{241}\mathrm{Am}$ spectrum with the $59.5~\mathrm{keV}$ full-energy peak and the simultaneously recorded pulser peak displayed at $102.31~\mathrm{keV}$. Panel (b) shows the $^{137}\mathrm{Cs}$ spectrum with the $662~\mathrm{keV}$ full-energy peak and the simultaneously recorded pulser peak displayed at $1030~\mathrm{keV}$. The y-axis is logarithmic. The fitted FWHM values labeled in the figure are used directly in the resolution analysis and in the derived spectroscopic metrics listed in Table~\ref{tab:gerc_spec_metrics}.}
  \label{fig:spec_kmrc01}
\end{figure}

KMRC01 spectra are presented in Fig.~\ref{fig:spec_kmrc01}. At $370~\mathrm{V}$ and the same shaping time of $0.5~\mu\mathrm{s}$, the $59.5~\mathrm{keV}$ line yielded a FWHM of $2.23~\mathrm{keV}$, while the simultaneous pulser width was $1.91~\mathrm{keV}$, as shown in Fig.~\ref{fig:spec_kmrc01}(a). The corresponding detector-related broadening term obtained after quadrature subtraction is $1.15~\mathrm{keV}$, indicating that the low-energy response is not purely electronics-limited. For the $^{137}\mathrm{Cs}$ spectrum in Fig.~\ref{fig:spec_kmrc01}(b), the $662~\mathrm{keV}$ photopeak width was $2.68~\mathrm{keV}$, only modestly larger than the pulser width of $2.43~\mathrm{keV}$, corresponding to a remaining detector-related term of $1.13~\mathrm{keV}$. Relative to SAP18, KMRC01 therefore exhibits substantially smaller non-electronic broadening at $662~\mathrm{keV}$ under the same shaping condition.

The present spectroscopy dataset is intentionally limited to two gamma-ray lines and a single shaping time of $0.5~\mu\mathrm{s}$ for each detector. This common setting was chosen to provide an initial, like-for-like comparison between the two first-generation prototypes rather than a full optimization of spectroscopic performance. As a result, the values reported here should be viewed as benchmark descriptors for compact process-validation devices. In particular, the current dataset does not separate ballistic-deficit effects from trapping, field nonuniformity, or contact-related broadening. A shaping-time scan and a broader multi-line calibration set will therefore be required in future work to determine whether the large detector-related term seen for SAP18 at $662~\mathrm{keV}$ is driven primarily by incomplete charge collection, depth-dependent collection times, or other geometry-specific effects.

To quantify the spectroscopic performance in a form that can be compared across energies, we use two complementary resolution metrics. The first is the conventional FWHM-based fractional energy resolution~\cite{Knoll2010_RDM,Haller1981_PhysicsUPGe,Bonet2021_EPJC_CONUS},
\begin{equation}
R(E)=100\times \frac{W_{\mathrm{tot}}(E)}{E},
\end{equation}
where $W_{\mathrm{tot}}(E)$ is the measured photopeak FWHM and $E$ is the gamma-ray energy, expressed in the same units. For Gaussian peak shapes, the corresponding standard deviation is~\cite{Knoll2010_RDM}
\begin{equation}
\sigma(E)=\frac{W_{\mathrm{tot}}(E)}{2.355}.
\end{equation}
The second metric follows the commonly used high-resolution gamma-spectroscopy parameterization in which the fractional width scales approximately as $1/\sqrt{E}$~\cite{Knoll2010_RDM,Haller1981_PhysicsUPGe,Mei2020_JPG_ChargeTrapping},
\begin{equation}
\frac{\sigma(E)}{E}=\frac{C(E)[\%]}{\sqrt{E[\mathrm{MeV}]}} ,
\end{equation}
which can be written equivalently as
\begin{equation}
C(E)=100\times \frac{\sigma(E)}{E}\sqrt{E[\mathrm{MeV}]}.
\end{equation}
In the present work, because only two full-energy peaks were measured for each prototype, $C(E)$ is quoted as a pointwise descriptor evaluated at each peak energy rather than as a global fit parameter extracted from a broad calibration set.

To estimate the detector-related broadening after removing the electronics contribution, the pulser width is subtracted in quadrature~\cite{Knoll2010_RDM,Jany2021_EPJC_Prototype,Bonet2021_EPJC_CONUS},
\begin{equation}
W_{\mathrm{det}}(E)=\sqrt{W_{\mathrm{tot}}^{2}(E)-W_{\mathrm{pul}}^{2}(E)},
\end{equation}
where $W_{\mathrm{pul}}$ is the simultaneously measured pulser FWHM. This quantity is included to show the residual broadening beyond the electronics response, but it is not used to define an additional resolution coefficient.

Using the fitted FWHM values reported in Figs.~\ref{fig:spec_sap18} and \ref{fig:spec_kmrc01}, SAP18 gives $R(59.5~\mathrm{keV})=4.10\%$ and $R(662~\mathrm{keV})=0.654\%$, while KMRC01 gives $R(59.5~\mathrm{keV})=3.75\%$ and $R(662~\mathrm{keV})=0.405\%$. The corresponding pointwise coefficients are $C(59.5~\mathrm{keV})=0.425\%$ and $C(662~\mathrm{keV})=0.226\%$ for SAP18, and $0.388\%$ and $0.140\%$ for KMRC01. These values are summarized in Table~\ref{tab:gerc_spec_metrics}. The spectra therefore confirm that both compact GeRC prototypes form identifiable full-energy peaks at cryogenic temperature, while also showing that the final resolution remains sensitive to detector-specific broadening mechanisms beyond the front-end electronics.

\begin{table}[t]
  \centering
  \caption{Summary of representative spectroscopic metrics for the compact GeRC prototypes at $77~\mathrm{K}$. Here $W_{\mathrm{tot}}$ is the measured photopeak FWHM, $W_{\mathrm{pul}}$ is the simultaneously recorded pulser FWHM, and $W_{\mathrm{det}}=\sqrt{W_{\mathrm{tot}}^{2}-W_{\mathrm{pul}}^{2}}$ represents the remaining detector-related broadening after subtraction of the electronics term in quadrature. The conventional fractional resolution is defined as $R=100\,W_{\mathrm{tot}}/E$. The coefficient $C(E)$ is calculated point-by-point from $\sigma/E=C(E)[\%]/\sqrt{E[\mathrm{MeV}]}$ with $\sigma=W_{\mathrm{tot}}/2.355$. Because only two full-energy peaks were measured for each detector, the quoted $C(E)$ values are pointwise descriptors at the indicated energies rather than global fit parameters over an extended calibration range.}
  \label{tab:gerc_spec_metrics}
  \small
  \resizebox{\textwidth}{!}{%
  \begin{tabular}{lcccccccc}
    \hline
    Prototype & $V_{\mathrm{op}}$ (V) & $\tau$ ($\mu\mathrm{s}$) & Peak & $W_{\mathrm{tot}}$ (keV) & $W_{\mathrm{pul}}$ (keV) & $W_{\mathrm{det}}$ (keV) & $R$ (\%) & $C(E)$ (\%) \\
    \hline
    SAP18  & 380 & 0.5 & $^{241}$Am, $59.5~\mathrm{keV}$ & 2.44 & 2.36 & 0.62 & 4.10  & 0.425 \\
    SAP18  & 380 & 0.5 & $^{137}$Cs, $662~\mathrm{keV}$  & 4.33 & 2.56 & 3.49 & 0.654 & 0.226 \\
    KMRC01 & 370 & 0.5 & $^{241}$Am, $59.5~\mathrm{keV}$ & 2.23 & 1.91 & 1.15 & 3.75  & 0.388 \\
    KMRC01 & 370 & 0.5 & $^{137}$Cs, $662~\mathrm{keV}$  & 2.68 & 2.43 & 1.13 & 0.405 & 0.140 \\
    \hline
  \end{tabular}%
  }
\end{table}

\paragraph{Interpretation and implications for process evolution.}
Although both prototypes function as working GeRC detectors, the overall performance is not yet optimized. Two factors are dominant in this first-generation study. First, the operational bias was intentionally kept in the few-hundred-volt regime, close to the depletion threshold, which reduces the field margin against low-field pockets and increases sensitivity to geometry- and contact-induced nonuniformities~\cite{Haller1981_PhysicsUPGe,Mertens2019_NIMA_LowNetImpurity,Comellato2021_EPJC_Collective}. Second, using sputtered a-Ge/Al thin films as the large-area HV electrode on a core-and-groove topology places stringent requirements on conformality and film quality in recessed regions (inside the bore and around the groove wings)~\cite{Thornton1974_JVST_SputterGeometry,Hull2005_aGeContacts,Wei2019_JINST_aGeDetectors}. Any local thinning, pinholes, or defect-mediated injection can increase leakage current and constrain the maximum bias, thereby indirectly impacting the achievable spectroscopic resolution~\cite{Hull2005_aGeContacts,Wei2020_EPJC_CBH,Panth2022_NIMA_TempBarrier}. The elevated leakage in KMRC01 is consistent with an additional defect-driven contribution from the observed grain-boundary features.

Importantly, these prototypes were fabricated primarily to demonstrate the feasibility of the most GeRC-specific processing steps, especially precision machining and prolonged polishing of non-planar surfaces. They were therefore not based on the highest-purity crystals available, and they should not be interpreted as direct surrogates for a mature large-mass p-type GeRC detector. With the mechanical and thin-film workflow now validated, the next milestone is to transition to higher-quality \emph{p}-type material and to implement a lithium-based outer electrode for robust high-voltage operation, using the planar hybrid-contact vehicle as the process reference~\cite{Morgan1969_GeLiPlanar,Yang2022_WraparoundLiAGe,Dong2026_arXiv_HybridProcess}. This step is expected to expand the bias margin substantially and to provide a more reliable path toward the ultimate GeRC objective: scalable, low-capacitance, point-contact-like signal formation in a ring-and-groove geometry suitable for next-generation arrays~\cite{Luke1989_IEEE_ShapedField,Barbeau2007_JCAP_PPC,Agostini2021_EPJC_ICPC}.

\section{Conclusions and Outlook}

This work demonstrates the first successful fabrication and cryogenic operation of compact all-thin-film GeRC prototypes. An end-to-end process flow was established for the geometry-specific steps that have limited prior GeRC development, including core-and-groove machining, non-planar surface conditioning, conformal a-Ge encapsulation, Al patterning, and GeRC-specific cryogenic mounting and readout. Both prototypes showed an inferred depletion onset near \(340~\mathrm{V}\), consistent with the electrostatic modeling, and both produced identifiable full-energy peaks from \(^{241}\mathrm{Am}\) and \(^{137}\mathrm{Cs}\) at 77 K. These results show that the recessed ring-contact topology can be realized as an operating HPGe detector geometry rather than remaining only a simulated concept.

At the same time, the present devices should be interpreted as first-generation process-validation prototypes rather than optimized spectroscopic detectors. The primary objective of this study was to verify that the core-and-groove machining and polishing sequence does not preclude stable low-temperature operation, and that conformal thin-film contacts can function on the non-planar GeRC surface prior to the introduction of lithium-diffused contacts in later device generations~\cite{Luke1992_aGeBipolar,Hull2005_aGeContacts,Wei2019_JINST_aGeDetectors,Morgan1969_GeLiPlanar,Yang2022_WraparoundLiAGe,Dong2026_arXiv_HybridProcess}. The observed device-to-device difference in leakage behavior, together with the detector-related peak broadening seen under the present operating conditions, indicates that contact quality, surface condition, and field uniformity on recessed GeRC features remain important limitations of the all-thin-film implementation used here~\cite{Haller1981_PhysicsUPGe,Hull2005_aGeContacts,Wei2020_EPJC_CBH,Panth2022_NIMA_TempBarrier,Mertens2019_NIMA_LowNetImpurity}.

The immediate next step is therefore to integrate the validated GeRC machining workflow with a lithium-based outer contact. As discussed in this work, that direction is motivated by the need for a more robust high-voltage electrode on the ring-and-groove topology~\cite{Morgan1969_GeLiPlanar,Yang2022_WraparoundLiAGe,Ghosh2025_LiGeContact,Dong2026_arXiv_HybridProcess}. The near-term priorities are to demonstrate conformal lithium deposition and controlled diffusion on recessed GeRC surfaces, to transfer the process to higher-quality detector crystals, and to re-evaluate leakage current, operating margin, charge-collection uniformity, and spectral response in that configuration. Beyond that stage, broader studies of dead-layer control, pulse-shape behavior, and fabrication reproducibility across multiple devices will be required before the scalability of the GeRC concept can be assessed realistically~\cite{Aguayo2013_NIMA_LiNplus,Andreotti2014_ARI_DeadLayer,Ma2017_ARI_InactiveLayer,Dai2022_arXiv_LiInactive,Agostini2013_GERDA_PSD,Alvis2019_MJD_MSE,Agostini2022_EPJC_GERDAPSA}.

More broadly, the significance of the present study lies in reducing a long-standing fabrication barrier in a detector concept aimed at future large-mass HPGe instrumentation. The motivation for GeRC is not simply geometric novelty, but the possibility of extending point-contact-like low-capacitance signal formation to larger single-crystal masses, where fewer detectors could in principle reduce channel count, cabling complexity, feedthrough burden, and passive surface area in ultra-low-background arrays~\cite{Luke1989_IEEE_ShapedField,Barbeau2007_JCAP_PPC,Cooper2012_NIMA_HPGeTrackingImaging,Agostini2021_EPJC_ICPC,Abgrall2021_LEGEND1000}. In that sense, this work does not yet establish a deployment-ready technology for tonne-scale experiments. It does, however, provide the first experimental process foundation for evaluating whether GeRC can mature into a practical complement to ICPC-style detectors in future rare-event searches that rely on large HPGe arrays~\cite{Ackermann2013_EPJC_GERDA,Arnquist2023_MJD_Final,Agostini2021_EPJC_ICPC,Abgrall2021_LEGEND1000}. If the remaining contact-integration and performance issues can be resolved in later generations, the GeRC approach may offer a credible path toward higher mass per channel while preserving the low-noise and pulse-shape-discrimination advantages that make HPGe detectors so powerful for next-generation low-background experiments~\cite{Luke1989_IEEE_ShapedField,Barbeau2007_JCAP_PPC,Agostini2013_GERDA_PSD,Alvis2019_MJD_MSE,Agostini2022_EPJC_GERDAPSA,Agostini2021_EPJC_ICPC}.

\acknowledgments

This work was supported in part by NSF OISE-1743790, NSF PHYS-2117774, NSF OIA-2427805,
NSF PHYS-2310027, and NSF OIA-2437416; by the U.S.\ Department of Energy under awards
DE-SC0024519 and DE-SC0004768; by the U.S.\ Air Force Office of Scientific Research under award
FA9550-23-1-0495; and by a research center supported by the State of South Dakota.
We thank Mark Amman for valuable technical advice on detector fabrication, and David C.~Radford and Felix Hagemann for helpful guidance and discussions related to the simulations.
We acknowledge Lawrence Berkeley National Laboratory for providing the cryostat used for detector characterization in this work.

\bibliographystyle{unsrt}
\bibliography{refs.bib}

\end{document}